\documentclass[aps,twocolumn,showpacs,pra]{revtex4}

\usepackage{epsfig}
\usepackage{amsmath}
\usepackage[latin1]{inputenc}
\usepackage{graphicx}

\newcommand{\nc}{\newcommand}
 \nc{\tcb}{\textcolor{blue}}  
 \nc{\tcr}{\textcolor{red}}
 \nc{\be}{\begin{equation}} 
 \nc{\ee}{\end{equation}}
 \nc{\bea}{\begin{eqnarray}}  
 \nc{\eea}{\end{eqnarray}}
 \nc{\ba}{\begin{array}}  
 \nc{\ea}{\end{array}}
 \nc{\rds}{{\rm d}s} 
 \nc{\rdt}{{\rm d}t} 
 \nc{\rdr}{{\rm d}r}
 \nc{\rdO}{{\rm d}\Omega} 
 \nc{\s}{{\rm S}} 
 \nc{\Pl}{{\rm Planck}}
 \nc{\dis}{\displaystyle} 
 \nc{\crit}{_{\rm cr}} 
 \nc{\rd}{{\rm d}}
 \nc{\munu}{{\mu\nu}} 
 \nc{\erm}{{\rm e}}
 \nc{\drm}{{\rm d}}
 \nc{\ov}{\overline}

\begin{document}

\title{Physical realization of Photonic Klein Tunneling}

\author{S. Esposito}
\email{sesposit@na.infn.it}%
\affiliation{
%\mbox{Dipartimento di Scienze Fisiche, Universit\`{a} di Napoli ``Federico II''} and 
Istituto Nazionale di Fisica Nucleare, Sezione di Napoli, Complesso Universitario di Monte
S.\,Angelo, via Cinthia, I-80126 Naples, Italy}

\

\

\begin{abstract}
\noindent  General physical conditions for the occurrence of photonic Klein tunneling are studied, where (controlled) spontaneous emission from the devices considered plays a key role. The specific example of a simple dielectric slab bounded by two dielectric half spaces with arbitrary refractive indices is worked out quite in detail, the measured reflection and transmission probabilities being calculated analytically. It is found that, in given cases, the measured reflection probability may be arbitrarily large (for large incident wavelengths) irrespective of the fact that the transmission probability is exponentially suppressed or not. Other interesting features of photonic Klein tunneling driven by (controlled) spontaneous emission are as well envisaged for practical applications.


\end{abstract}

%78.67.Pt 	Multilayers; superlattices; photonic structures; metamaterials
%Photonic band gap materials, 42.70.Qs
%Tunneling
%quantum mechanics of, 03.65.Xp, 03.75.Lm

\maketitle

\noindent The experimental observation \cite{expobs} of the intriguing properties of graphene \cite{graphrev}, a genuine two-dimensional material composed of carbon atoms forming a honeycomb lattice, has recently drawn a very vivid interest by many physicists (see \cite{graphrev}, \cite{graphDirac} and references therein). The main reason relies in the fact that the forbidden energy gap, through which the charge carriers tunnel, is suppressed to zero in graphene, differently from other materials, and, in addition, those peculiar charge carriers behave as Dirac fermions \cite{graphDirac}, \cite{KNG}. This means that electrons in such unique material behave effectively as massless Dirac fermions, thus allowing the possible experimental study of a number of key phenomena \cite{key}. Indeed, the scattering of relativistic fermions, being described by the Dirac equation, is fundamentally different from that of non-relativistic ones. For example, relativistic electrons scattered off a potential step, with height higher than their energy, at normal incidence may exhibit a non-zero transmission probability, this being in sharp contrast to the intuitive result of non-relativistic quantum mechanics. Such a phenomenon, in general referred to as Klein tunneling \cite{Calogeracos}, that is the below-barrier particle tunneling without the exponential damping expected for non-relativistic particles, has never been observed experimentally, since too much strong electric fields are required to observe it in elementary particles. Nevertheless, Klein tunneling may be realized on graphene sheets \cite{KNG}: since a sufficiently strong potential, though repulsive for electrons, is attractive for holes, it gives rise to hole states in the barrier to form channels through which electrons can penetrate the barrier \cite{Calogeracos}. Several experiments have been made to observe such phenomenon, and we refer the interested reader to the existing literature \cite{grapheneexp} for further discussion.

Very recently, even in view of possible intriguing applications, optical analogues of Klein tunneling have been as well proposed \cite{opticalKT}-\cite{Longhi}, thus capturing the interest of many. This is the case, for example, of light propagation in deformed honeycomb photonic lattices \cite{Bahat} \cite{LonghiB81}, whose band structure is similar to that of graphene, or light refraction at the interfaces between positive and negative index materials \cite{Duney}. Other proposals regard spatial light propagation in binary waveguide arrays \cite{LonghiB81}, one-dimensional stationary light pulses in an atomic ensemble with electromagnetic induced transparency (in the limit of tight spatial confinement) \cite{Otterbach}, or Klein tunneling of light in fiber Bragg gratings \cite{Longhi}.

In these papers, the investigations on photonic analogues of relativistic tunneling phenomena are carried out exactly as made for analogues of non-relativistic tunneling (see, for example, \cite{Esposito} and references therein), that is by writing down the dynamical equations for the light propagating in the device considered and then exploiting the formal mathematical analogy between such equations and the Dirac equation for electrons. For example, in \cite{Longhi} it is shown that the counter-propagating waves in a fiber Bragg grating satisfy coupled-mode equations that can be exactly cast in the form of a Dirac equation in presence of an electrostatic field, provided a suitable definition for a ``spinor" wavefunction. The optical analogue of the forbidden energy region is given by the photonic stop band of the periodic grating. Pulse propagation in a fiber Bragg grating with a suitably designed chirp profile can then be used to mimic the relativistic tunneling of a wavepacket in a potential step. Other, different realizations, as mentioned above, are treated analogously.

Such a way of reasoning is certainly legitimate, and leads to correct results. Nevertheless, by focusing mainly on the formal analogy among mathematical equations does not provide a full physical insight in the true phenomenon considered, and thus only particular cases are usually studied. In this note, instead, we will discuss some general conditions for the realization of photonic Klein tunneling, mainly focusing on the physical interpretation of it.

\begin{figure}
\begin{center}
\epsfig{height=2.5truecm,file=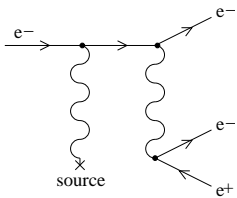} 
\caption{A diagrammatic representation of Klein tunneling of electrons by an electrostatic potential source.}
\label{fig1}
\end{center}
\end{figure}

\begin{figure}
\begin{center}
\begin{tabular}{cl}
\epsfig{height=3.3truecm,file=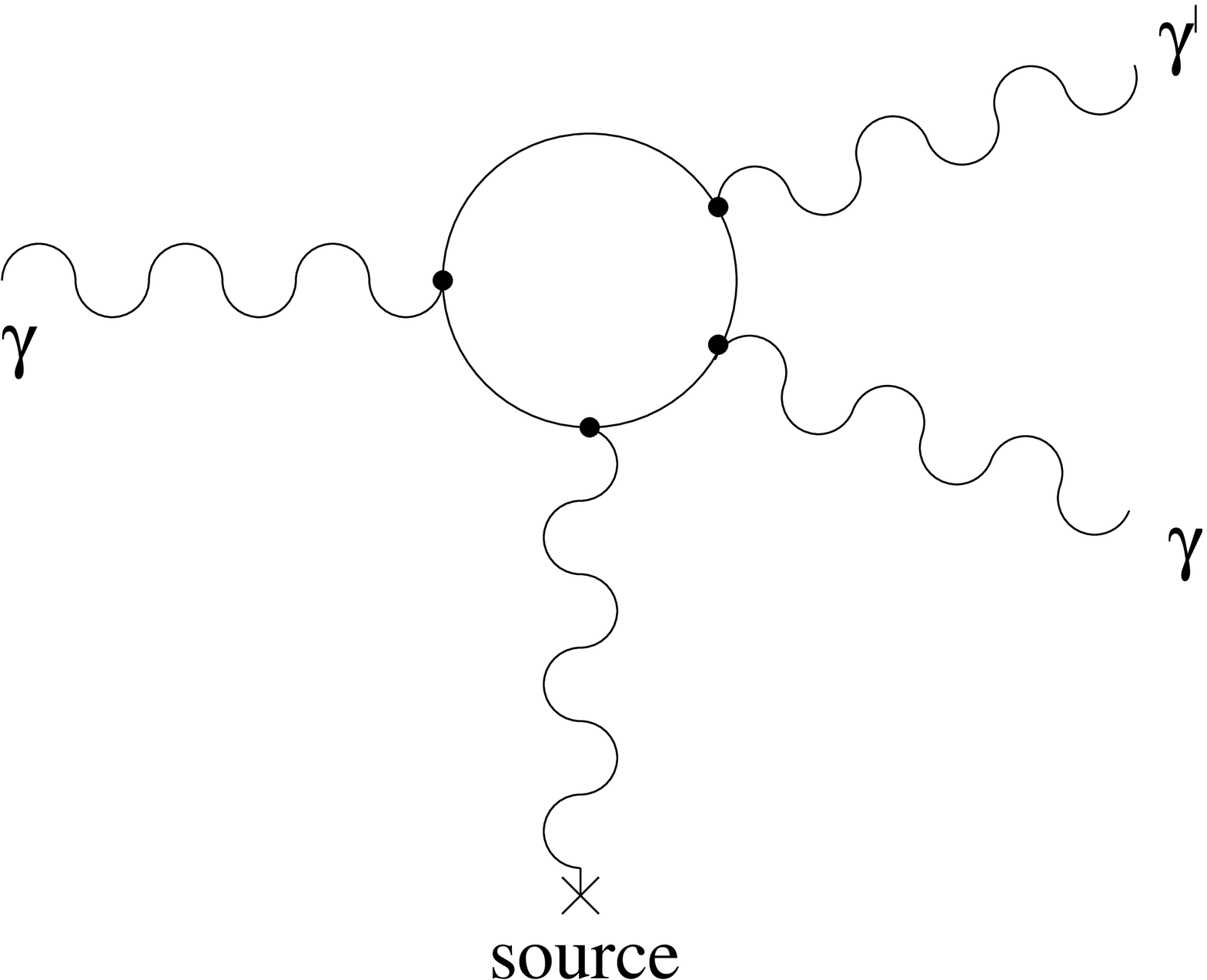} & a) \\ \\
\epsfig{height=2truecm,file=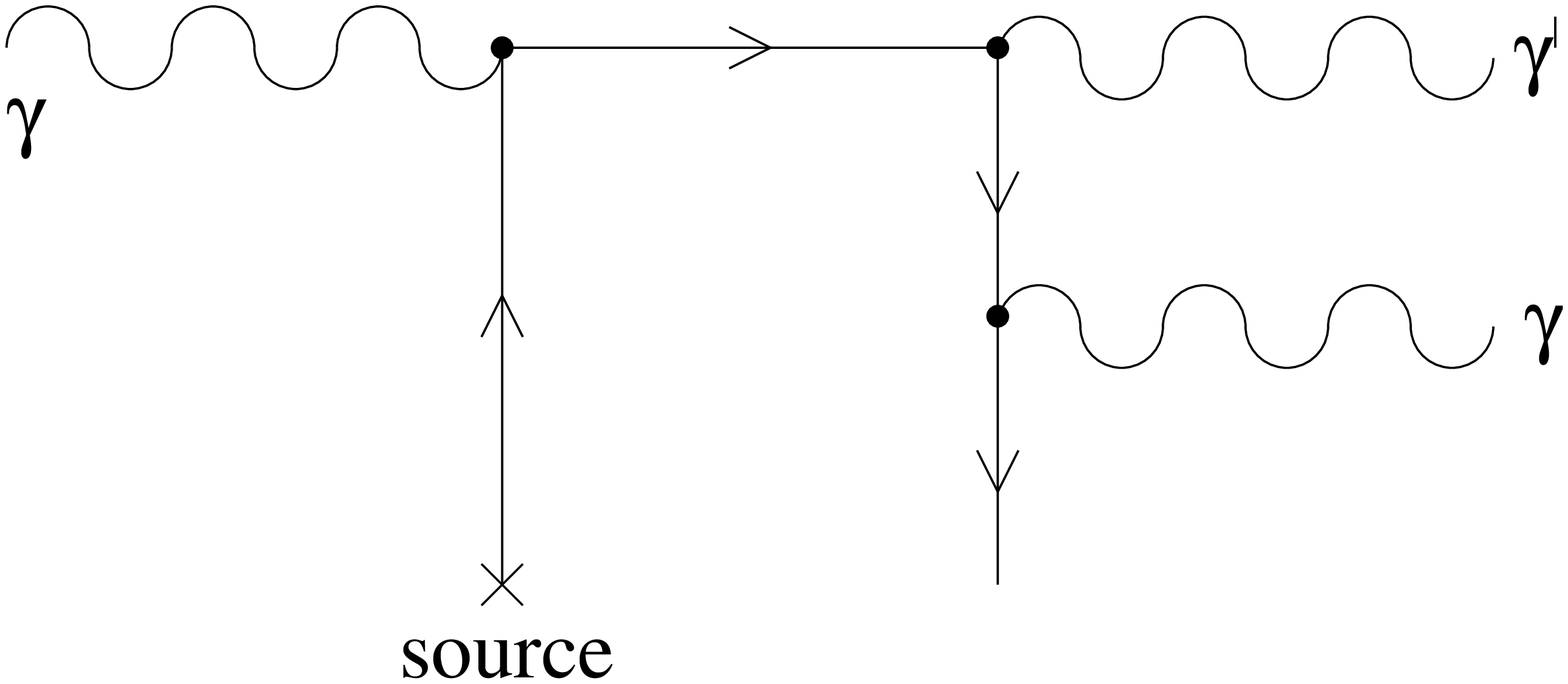} & b)
\end{tabular}
\caption{A diagrammatic representation of photonic Klein tunneling: a) by an electromagnetic potential source; b) by an electron source (material medium).}
\label{fig2}
\end{center}
\end{figure}

Let us start with the typical example of an electron (with relativistic energy) impinging on a potential barrier of height $V_0$ created, for example, by a sufficiently strong electric field. As well known, Klein tunneling takes place only when $V_0$ is larger than twice the particle rest energy $2 m_e c^2$, so that pair production can occur, according to the diagrammatic representation \footnote{Recall, however, that tunneling is a non-perturbative phenomenon, so that it cannot be described by perturbative Feynman diagrams. This and the following figures have, then, only illustrative purposes.} in Fig. \ref{fig1}. For photonic tunneling, such a limitation is not present, since photons have no rest energy, so that, in principle, photonic Klein tunneling should occur more easily. This simple expectation, however, confronts with the effective realization of the potential barrier for the photons or, in other words, with the coupling of the incoming photon to the barrier: for electron tunneling, indeed, electrons couple directly to the electric field originating the potential barrier. A naive transposition of electron tunneling to photonic tunneling would proceed according to the diagrammatic representation in Fig. \ref{fig2}a, where photon-photon coupling even in vacuum is indeed possible in a full quantum picture (at least at one loop level) through the creation of virtual electron-positron pairs, when non-linear effects come out. In such a case (photons impinging on an ``electromagnetic barrier"), Klein tunneling effectively reduces to the phenomenon of photon splitting, extensively considered in the literature \cite{splitting}. 
The addition of non-linear terms to the wave equation (in the Euler-Heisenberg approximation) \cite{NLMaxwell},
\bea 
\left( \nabla^2 - \frac{1}{c^2} \, \frac{\partial^2 }{\partial t^2} \right) {\bf E} &=& \mu_0 \left[
\frac{\partial^2 {\bf P}}{\partial t^2} + c^2 \nabla \left( \nabla \cdot {\bf P} \right)  \right. \nonumber  \\
& & \left. \;\; \;\; \;\; \;\;  \;\; \;\;  \;\;  \;\; \;+ \frac{\partial }{\partial t} \left( {\mathbf \nabla} \times {\bf M} \right) \right]
\label{1}
\eea
(and similarly for the ${\bf B}$ field), where the ``background" (source) quantities are given by
\bea
{\bf P} &=& 2 \zeta \left[ 2 \left( E_s^2 - c^2 B_s^2 \right) {\bf E}_s + 7 c^2 \left( {\bf E}_s \cdot 
{\bf B}_s \right) {\bf B}_s \right] \label{2} \\
{\bf M} &=& - 2 c^2 \zeta \left[ 2 \left( E_s^2 - c^2 B_s^2 \right) {\bf B}_s + 7 \left( {\bf E}_s \cdot {\bf B}_s \right) {\bf E}_s \right] \label{3}
\eea
with $\zeta = 2 \alpha^2 \epsilon_0^2 \hbar^3 / 45 m_e^4 c^5$ ($\alpha = e^2/2 \epsilon h c$ is the fine structure constant), under given conditions allow the photon propagation in a way similar to that in an undersized waveguide, thus mimicking a tunneling phenomenon. However, as largely debated in the literature \cite{splitting} \cite{NLMaxwell}, and as evident from the equations above, such effect is almost unobservable in laboratory experiments and certainly not of practical use in appliction, due mainly to the extremely small value of the effective coupling constant $\zeta$, while it may play a non-negligible role in several astrophysical environments. For this reason, we do not consider here such a possibility.

An alternative comes from the coupling of the tunneling photon to an ``electron" (rather than electromagnetic) source, as diagrammatically depicted in Fig. \ref{fig2}b; that is, the potential barrier is given by electrons in a suitable material, such as a refractive medium. This is not an example strictly equivalent to electron tunneling, but is a common and easy way to achieve (even standard, not Klein) photonic tunneling. In such a case, referring to Fig. \ref{fig2}b, the additional photon \footnote{Here and above, by contrast to the electron tunneling case, it is not necessary to have the production of photon {\it pairs}; however we limit ourselves to the most probable situation of only one additional photon.} is produced by the electrons in a given medium by means of spontaneous emission, which is, again, a pure quantum effect. The physical realization of this version of photonic Klein tunneling may proceed through different mechanisms, but its practical achievement is related to the possibility of effective (and controlled) spontaneous emission.

As an example, we may consider the spontaneous emission of photons by impurity atoms or molecules in homogeneous dielectrics. The vacuum emission rate of a randomly oriented dipole with moment $d$ is given by \cite{Jackson}
\be  \label{4}
\Gamma_0 = \frac{d^2 \omega_0^3}{3 \pi \epsilon_0 \hbar c^3}
\ee
for a transition of frequency $\omega_0$, but, as it is now well established, the rate at which an excited species undergoes spontaneous emission depends on its environment \cite{mirrorcavities}
\cite{SEDielectrics}. The spontaneous emission rate in a bulk dielectric, for instance, is scaled by the real part of the refractive index $n =\sqrt{\epsilon}$ of the medium at the frequency of the transition \cite{SEDielectrics}. In general, the spontaneous emission field can be strongly modified if the emitting atom is surrounded by materials of different composition and shape, since the existence of material boundaries (of any kind) in the vicinity of the radiating species simply changes the strength and distribution of the electromagnetic modes with which the emitter interacts, then resulting in an altered spontaneous emission rate. In particular, when an increase in the density of states occurs, the rate of spontaneous emission can be enhanced over the free space value (\ref{4}). Similar results are obtained when the atoms are placed between mirrors or in resonant cavities \cite{mirrorcavities}, in which case larger modifications are possible. 

Spontaneous emission rates are also altered in photonic crystals \cite{photoniccrystals}, where the dielectric constant is periodically modulated on length scales of the wavelength of light. Due to multiple Bragg reflections, light in a certain frequency range cannot propagate in the crystal, so that the radiative density of states in this photonic band gap is zero and, therefore, spontaneous emission of excited atoms embedded in these materials is expected to be inhibited. However, at other frequencies it has been shown that spontaneous emission can be either enhanced or reduced, so that in photonic band gap materials a nearly complete control over spontaneous emission rates might be achieved. This opens the intriguing possibility to have a single material where standard and Klein tunneling may be observed for incident photons with different frequencies.

A number of other devices (left-handed materials \cite{others1}, sharp metallic tips \cite{others2}, waveguides \cite{others3}, quantum dots \cite{others4}, etc.) may as well be used to modify the emission features, all of them driven by the zero-point fluctuations of the electromagnetic field. It is at last clear that the key ingredient relevant for the practical realization of the photonic Klein tunneling we are dealing with is just the general result that spontaneous emission of an emitter can be controlled to some extent, and then engineered by tailoring the surrounding structure on a transition wavelength scale.

\begin{figure}
\begin{center}
\epsfig{height=4truecm,file=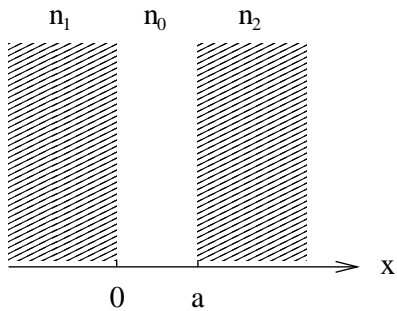} 
\caption{Photonic Klein tunneling through a dielectric slab.}
\label{fig4}
\end{center}
\end{figure}

Let us now address the question of finding general conditions for photonic Klein tunneling to occur. In order to get definite results, we shall consider a specific example for illustrative purposes, that is a dielectric slab bounded by two dielectric half spaces with arbitrary refractive indices, as shown in Fig. \ref{fig4}, with the incident wave coming from the negative $x$ region. The photonic propagation may be described simply by a scalar field, which we denote with $\psi$ in analogy to the Dirac case: it corresponds to some scalar component of the electric or magnetic field satisfying the appropriate electromagnetic Helmholtz equation. In the most general case, by allowing non-vanishing spontaneous emission in the two barrier-free regions (in both negative and positive $x$ directions), the solution of such equation can be written as
\be
\psi(x) = \left\{ \ba{ll}
\psi_1(x) , & \;\;\; {\rm for} \; x \leq 0, \\
\psi_0(x) , & \;\;\; {\rm for} \; 0 \leq x \leq a, \\
\psi_2(x) , & \;\;\; {\rm for} \; x \geq a, 
\ea \right.
\ee
\noindent with
\be
\ba{l}
\dis \psi_1(x) = \left[ 1 + S_1(x) \right] \erm^{i k_1 x} \\ 
\;\;\; + \left\{ R \left[ 1 +  S_1(x) \right] + S_1(x) + T \, S_2(x) \right\} \erm^{-i k_1 x} , 
\\  \\
\dis \psi_0(x) = A \, \erm^{- \chi_0 x} + B \, \erm^{\chi_0 x} ,  \\ \\
\dis \psi_2(x) = S_2(x) \, \erm^{-i k_2 (x-a)} \\ 
\;\;\; + \dis \left\{ T \left[ 1 + S_1(x) \right] + S_2(x) + R \, S_2(x) \right\} \erm^{i k_2 (x-a)} ,
\ea
\label{e1} 
\ee
where $k_1, k_2$ are the real wavevectors in the barrier-free regions, $i \chi_0$ is the imaginary wavevector in the $0 \leq x \leq a$ region, and $R,T$ are the reflection and transmission coefficient of the barrier, respectively. The real source functions $S_1(x), S_2(x)$ take into account the spontaneous emission rates (and, if any, possible absorptive effects): for a dielectric slab as that considered here, these have been calculated in \cite{Rikken} as functions of the position, so that we refer the reader to this paper for further discussion. However, since we are interested in general conditions, the explicit (though numerical) expressions for these are not really necessary here, as we shall see below. Note that, in $\psi_1(x)$, the $S_1(x)$-terms accounts for the (isotropic) spontaneous emission in region 1, the $R \, S_1(x)$-term accounts for the reflection of such signal by the barrier, while the $T \, S_2(x)$-term accounts for the spontaneous emission signal in region 2 transmitted by the barrier in region 1. Analogously, in $\psi_2(x)$, the $S_2(x)$-terms accounts for the (isotropic) spontaneous emission in region 2, the $R \, S_2(x)$-term accounts for the reflection of such by the barrier, while the $T \, S_1(x)$-term accounts for the spontaneous emission signal in region 1 transmitted by the barrier in region 2. \\
The matching conditions at interfaces are given  by
\be \label{e2}
\ba{rcl}
\dis \psi_1(0) = \psi_0(0) , & & \dis \psi^\prime_1(0) = \psi^\prime_0(0), \\ 
\dis \psi_0(a) = \psi_2(a) , & & \dis \psi^\prime_0(a) = \psi^\prime_2(a), 
\ea
\ee
where the prime denotes differentiation with respect to $x$. By substitution into Eq. (\ref{e1}) we can find the explicit expressions for the function $\psi$ and, then, for the reflection and transmission probabilities. However, the resulting expression for such general case are excessively lengthy and no useful physical insight can be gained easily. We instead consider few interesting special cases, relevant for practical applications, the physical meaning of the corresponding results being particularly transparent. In addition, we consider a symmetric barrier, with $k_2=k_1$ (for $k_2 \neq k_1$ the calculations are a bit more involved, while no further physics comes out), in the fully opaque limit $\chi_0 a \rightarrow \infty$. The last assumption is justified when focusing on the dominant terms of the Klein tunneling, without the characteristic exponentially decreasing factors $\erm^{- 2 \chi_0 a}$. Of course, corrections including such factors are always present, but here we neglect their effect by considering only very opaque barriers. \\
{\it First case.} Let us assume that spontaneous emission takes place only in region 1, and only along the positive $x$ direction, as for the incident signal. In such a case, $\psi_1(x)$ and 
$\psi_2(x)$ in Eqs. (\ref{e1}) reduce to:
\be
\ba{l}
\dis \psi_1(x) = \left[ 1 + S_1(x) \right] \erm^{i k_1 x} + R \left[ 1 +  S_1(x) \right]  \erm^{-i k_1 x} , 
\\ 
\dis \psi_2(x) = T \left[ 1 + S_1(a-x) \right]  \erm^{i k_1 (x-a)} .
\ea
\label{case1} 
\ee
We have, then, just a rescaled $\psi$ function in the barrier-free regions with respect to the standard case (with no spontaneous emission). As expected, by imposing the matching conditions (\ref{e2}), in the approximations above we have the following results for the reflection and transmission probabilities by the barrier:
\be
\left| R \right|^2 \simeq 1 , \qquad \left| T \right|^2 \simeq 0.
\label{r1}
\ee
However, these probabilities are not directly obtained in experiments, since the measured probabilities are normalized with respect to {\it only} the signal emitted by the source employed and not to the total signal impinging on the barrier (including the spontaneous emission source). In the present case, then, the measured reflection and transmission probabilities are evaluated from the ratios $\left| R \right|^2 \left[ 1 + S_1(x) \right]^2$ and $\left| T \right|^2 \left[ 1 + S_1(a-x) \right]^2$, respectively. Given the effective $x$ dependence, these ratios should be integrated over the effective measurement regions but, taking into account the fact that the spontaneous emission rate is dominant near the edges of the barrier \cite{Rikken}, as a good approximation we can evaluate the source terms above just at the corresponding edges \footnote{That is, at $x=0$ for the reflection probability and at $x=a$ for the transmission one.}, thus obtaining
\be 
\left| R_{\rm measured} \right|^2 \simeq \left[ 1 + S_1(0) \right]^2 , \qquad
\left| T_{\rm measured} \right|^2 \simeq 0 .
\label{reff1}
\ee
From these results we get the quite obvious conclusion that, in the present case and in the approximations above, the reflection probability is {\it greater} than one due to the spontaneous emission, while the transmission probability is exponentially suppressed. \\
{\it Second case.} Let us now assume that spontaneous emission takes place again only in region 1, but only along the negative $x$ direction, as for the incident signal reflected by the barrier. In such a case, $\psi_1(x)$ and $\psi_2(x)$ in Eqs. (\ref{e1}) reduce to:
\be
\ba{l}
\dis \psi_1(x) = \erm^{i k_1 x} + \left[ R +  S_1(x) \right]  \erm^{-i k_1 x} , 
\\ 
\dis \psi_2(x) = T \, \erm^{i k_1 (x-a)} .
\ea
\label{case2} 
\ee
In the same approximations as above, the measured reflection and transmission probabilities are now given by $\left| R  + S_1(0) \right|^2$ and $\left| T \right|^2$ respectively, so that, by imposing the matching conditions (\ref{e2}), we get:
\be 
\ba{l}
\dis \left| R_{\rm measured} \right|^2 \simeq \frac{k_1^2 + \left[ \chi_0 + S^\prime_1(0) \right]^2}{k_1^2 + \chi_0^2} , \\
\dis \left| T_{\rm measured} \right|^2 \simeq 0 .
\ea
\label{reff2}
\ee
Again, then, signal transmission is exponentially suppressed, but now the reflection probability exhibits an interesting property: it is {\it greater} than one if $S^\prime_1(0)$ is positive, while it keeps {\it lower} than one in the opposite case. In other words, the super-unitary reflection is determined by the sign of the slope of the spontaneous emission rate which, as deduced from the results reported in Ref. \cite{Rikken}, depends on the characteristics of the apparatus employed (refractive indices of the slab, emission wavelength and thickness of the slab). \\
{\it Third case.} The two particular cases just considered are generalized by assuming that the spontaneous emission (taking place only in region 1) is isotropic, that is it takes place along both the positive and negative $x$ directions. Now, then, $\psi_1(x)$ and $\psi_2(x)$ in Eqs. (\ref{e1}) reduce to:
\be
\ba{l}
\dis \psi_1(x) = \left[ 1 +  S_1(x) \right] \erm^{i k_1 x} \\
\qquad \qquad + \left\{ R \left[ 1 +  S_1(x) \right]  +  S_1(x) \right\}  \erm^{-i k_1 x} , 
\\ 
\dis \psi_2(x) = T \left[ 1 +  S_1(a-x) \right]  \erm^{i k_1 (x-a)} .
\ea
\label{case3} 
\ee
The measured reflection and transmission probabilities are given by $\left| R \left[ 1 +  S_1(0) \right]  + S_1(0) \right|^2$ and $\left| T \left[ 1 +  S_1(0) \right] \right|^2$ respectively, and, by imposing the matching conditions (\ref{e2}), we have:
\be 
\ba{l}
\dis \left| R_{\rm measured} \right|^2 \simeq \frac{\ov{k}_1^2 + \left[ \ov{\chi}_0 + 
\ov{S}^\prime_1(0) \right]^2}{\ov{k}_1^2 + \ov{\chi}_0^2} \,  \left[ 1 +  S_1(0) \right] , \\
\dis \left| T_{\rm measured} \right|^2 \simeq 0 ,
\ea
\label{reff3}
\ee
where $\ov{k}_1 = k_1 [ 1 +  S_1(0)]$, $\ov{\chi}_0 = \chi_0 [ 1 +  S_1(0)] + S^\prime_1(0)$, 
$\ov{S}^\prime_1(0) = S^\prime_1(0) / [ 1 +  S_1(0)]$. Even in the present case, then, the super-unitary reflection depends on the sign of the slope of the source term; in particular, again the reflection probability is certainly {\it greater} than one for $S^\prime_1(0) >0$. \\
{\it Fourth case.} As a final example, let us consider the case with the spontaneous emission taking place only in region 2, along the negative $x$ direction (of course, there is no interesting result for the reflection probability when light is spontaneously emitted only along the positive $x$ direction, while changing the transmission properties in a fashion similar to that of the second case above). Now, $\psi_1(x)$ and $\psi_2(x)$ in Eqs. (\ref{e1}) reduce to:
\be
\ba{l}
\! \dis \psi_1(x) = \erm^{i k_1 x} + \left[ R +  T \, S_2(a-x) \right]  \erm^{-i k_1 x} , 
\\ 
\! \dis \psi_2(x) = \left[ T + R \, S_2(x) \right] \erm^{i k_1 (x-a)} + S_2(x) \, \erm^{i k_1 (x-a)} .
\ea
\label{case4} 
\ee
The measured reflection and transmission probabilities are given by $\left| R + T \, S_2(a) \right||^2$ and $\left| T + R \, S_2(a)  \right|^2$ respectively. Differently from the previous cases, the transmission probability is now {\it not} exponentially suppressed; indeed, by imposing the matching conditions (\ref{e2}), we get:
\be 
\ba{l}
\dis \left| R_{\rm measured} \right|^2 \simeq \frac{\left[ (k_1^2 + \chi_0^2) - \kappa^2 \right]^2 + 4 S_2^2(a) S_2^{\prime 2}(a) k_1^2}{\left[ (k_1^2 + \chi_0^2) - \kappa^2 \right]^2 + 4 S_2^{\prime 2}(a) k_1^2}, \\
\dis \left| T_{\rm measured} \right|^2 \simeq \frac{S_2^2(a) \left[ (k_1^2 + \chi_0^2) - \kappa^2 \right]^2 + 4 S_2^{\prime 2}(a) k_1^2}{\left[ (k_1^2 + \chi_0^2) - \kappa^2 \right]^2 + 4 S_2^{\prime 2}(a) k_1^2} ,
\ea
\label{reff4}
\ee
where $\kappa^2 = S_2^2(a) k_1^2 + \left[ S_2(a) \chi_0 - S_2^\prime(a) \right]^2$. Note that both the reflection and the transmission probabilities are {\it greater} than one (as expected) for {\it any}
value of $S_2(a)$ and value and sign of $S_2^\prime(a)$. The total probability is, of course, again greater than one and, differently from the second and third case above, does not depend explicitly on $k_1, \chi_0$, the following relation holding true:
\be
\left| R_{\rm measured} \right|^2 + \left| T_{\rm measured} \right|^2 \simeq 1 + S_2^2(a).
\label{one}
\ee
A more general case, which is a better representation of physical reality, would include spontaneous emission from both barrier-free regions but, in such a case, mathematical results become less transparent. Nevertheless, even in this general case the reflection probability is greater than one, while tunneling is not exponentially damped. 

\

From the results obtained above, it emerges quite clear that photonic Klein tunneling is not so much due to the mathematical structure of the electrodynamic equations, similar to the Dirac equation for electrons, but rather the physical content behind them. For electrons, this translates particle-antiparticle pair creation from the potential barrier. For photons, particle production from an electromagnetic potential barrier is as well a possible phenomenon due to effective non-linearities in the Maxwell equations arising from quantum loop corrections (the phenomenon reduces, in a sense, to the well-known photon splitting); however, the extremely smallness of such corrections renders the effect of no practical use in applications. Instead, general Klein tunneling of photons may be easily realized when they impinge on an ``electron'' (rather than electromagnetic) barrier, that is a refractive medium, and spontaneous photon emission takes place. This effect is of somewhat practical interest since spontaneous emission, contrary to naive expectations, can be controlled in such media (for example, in photonic band gap materials) to some extent, as recalled above. Interestingly enough, as shown in the specific example of a dielectric slab considered here, photonic Klein tunneling may be realized even independently of the features of the spontaneous emission sources (fourth case). With the exception of the first case considered above, where spontaneous emission takes place only in the incident signal region along the direction of the incident signal, the measured reflection probability depends on $k_1$ and $\chi_0$, and may be made {\it arbitrarily large} for large incident wavelengths (and small imaginary wavevector of the barrier medium). Equally interesting, likely even in applications, is the prediction that the measured reflection probability may be {\it lower} than that in the standard scenario (no Klein tunneling) if spontaneous emission takes place in the incident signal region along the direction opposite to that of the incident signal (second and third cases), for given features of the source (negative slope of the spontaneous emission rate).

The results obtained here for a simple dielectric slab may be generalized to different devices, so that a variety of possible implementations of photonic Klein tunneling may be realized in several applications, provided that the corresponding spontaneous emission properties can be easily managed. Such a challenge then claims for more theoretical and experimental studies, which will certainly lead to novel applicative solutions in the very near future.

\end{document}